\documentclass{svjour2}
\smartqed
\usepackage{amssymb,latexsym} 
\usepackage{mathptmx,mathtools,graphicx,amssymb,epsfig}
\usepackage{graphicx}
\usepackage[all, knot]{xy}
\xyoption{arc}

\journalname{Foundations of Physics}

\begin{document} 

\title{Mirror Symmetry and Other Miracles in Superstring Theory\footnote{I would like to thank audiences at the FQXi meeting in the Azores in 2009, and in Toronto, Oxford, Leeds, Utrecht, Bradford, and Sydney for discussions. I would also like to acknowledge the Australian Research Council for financial support.}}

\author{Dean Rickles\footnote{For submission to a \emph{Foundations of Physics} special issue on ``Forty Years Of String Theory: Reflecting On the Foundations'' (edited by G. `t Hooft, E. Verlinde, D. Dieks and S. de Haro).}}

\institute{Dr D. Rickles \at
               \\Unit for HPS, University of Sydney\\
              \email{d.rickles@usyd.edu.au}}

\date{Received: date / Accepted: date}

\maketitle

\begin{abstract}
The dominance of string theory in the research landscape of quantum gravity physics (despite any \emph{direct} experimental evidence) can, I think, be justified in a variety of ways. Here I focus on an argument from mathematical fertility, broadly similar to Hilary Putnam's `no miracles argument' that, I argue, many string theorists in fact espouse. String theory leads to many surprising, useful, and well-confirmed mathematical `predictions'---here I focus on mirror symmetry. These predictions are made on the basis of general \emph{physical} principles entering into string theory. The success of the mathematical predictions are then seen as evidence for framework that generated them. I attempt to defend this argument, but there are nonetheless some serious objections to be faced. These objections can only be evaded at a high (philosophical) price.

\keywords{String Theory \and Mirror symmetry \and No Miracles Argument}
\end{abstract}

\section{Experimental Distance in Quantum Gravity}

Like most (if not all) quantum gravity research, string theory is bound to increase the amount of indirectness between theory and experiment. The scale at which unique, quantitatively determinable new predictions are made is well beyond the reach of any experiment, past, present, or (conceivable) future. We should therefore \emph{expect} that new methods of theory evaluation will arise to compensate for this.\footnote{Of course, the standard collider methods in particle physics have an in-built (practical) termination point since deeper scales require ever higher energies.} Schr\"{o}dinger inadvertently pointed toward this issue in 1955:

\begin{quotation}
\noindent  It might [have been] the case that in experimental physics the method for establishing laws were the same as in astronomy. ... But it is not so. And that is small wonder. The physicist has full liberty to interfere with his object and to set the conditions of experiment at will. (\cite{schro}, p. 13)
\end{quotation} 

\noindent If the physicist loses the ability to interfere with his objects (as in string theory), then the implication would seem to be that the method of establishing laws and facts must thereby be modified. In this case, there would appear to be three broad strategies: 

\begin{enumerate}
\item Shift to the observational methods of precisely the kind relied upon by astronomers and cosmologists.
\item Reduce the emphasis placed on \emph{quantitative} predictions (in favour of weaker, qualitative predictions).
\item Attempt to utilise a range of other theoretical virtues, such as the ability of a theory to unify a broad range of disparate (old) knowledge.  
\end{enumerate}

\noindent String theory makes use of all three of these methods in varying degrees: along the lines of the first strategy, though still in an embryonic stage, string cosmology is emerging (in which the exceptionally large energies/small scales of the very early universe are utilised in a bid to find stringy remnants); in the second case one can point to supersymmetry, holography, and quantum geometry, in the former case, and the theoretical unification of gauge theory and gravity\footnote{Physicists often refer to these as `retrodictions', though philosophers refer to such instances as `accommodations'. I will discuss, in \S\ref{meth}, the issue of the relative weights assigned to prediction on the one hand and accommodation on the other, for it is a matter that divides philosophers of science (and statisticians)---see \cite{lipton}. Note, in any case, that philosophers and historians of science are usually suspicious of anything claiming to be `the scientific method': the notions of `testability' and `falsifiability' are, in particular, notoriously flawed. (It is somewhat shocking to see that in his review of Lee Smolin's \emph{The Trouble with Physics}, Michael Riordan, a Stanford University historian of science no less, claims that string theory is scientifically on a par with the theory of intelligent design (\cite{rio}, p. 39)! He sees science as tantamount to the production of testable predictions. This demonstrates a woeful ignorance of much of the painstaking work that historians and philosophers of science, since Pierre Duhem and, more obviously, Kuhn, have carried out.)} in the latter case. 

It is well-known, then, that string theory doesn't have an experimental leg to stand on, at least not by way of \emph{novel} experimentally testable predictions.\footnote{I'm referring to string theory \emph{qua} quantum theory of gravity (or TOE). There are several instances of string theoretic \emph{models} being usefully employed to make empirical predictions. For example in the study of quark-gluon plasmas in heavy ion physics (such as those produced at RHIC) \cite{myers}, and most recently in high $T_{c}$ superconductivity \cite{hand}. It is essentially the conformal invariance, coupled with the holographic principle  that does the work here---both of which are, of course, more general than superstring theory. However, since string theory implies these features, one could still make a case that they `weakly confirm' the framework of string theory.} And there are intuitively obvious reasons why this should be so, as a result of the scales involved in the new physics. It is also the case that string theory scores highly on its ability to unify `old evidence', and it is generally understood that this puzzle-solving ability is what gives string theory its credibility. Though there is certainly truth in this claim, in this paper I present an alternative account for the credibility of string theory, and argue that it is more likely than not the argument that underlies the faith of string theorists, and also mathematicians who study the theory. The argument I present bears striking similarity to the Smart-Putnam `no-miracles argument' from the philosophy of science literature \cite{putnam,smart}. The crucial difference is that the `miracles' are not surprising physical facts but surprising \emph{mathematical} facts instead. After presenting this mathematical version of the no-miracles argument I then attempt to defend it as offering support to string theory \emph{qua} physical theory. 

We begin with a brief tour of the curious history of string theory, highlighting the way in which string theory was born from general physical principles. These principles will play a crucial role in the subsequent argument. We then introduce the (Calabi-Yau) mirror symmetry example that forms the core of this argument. Finally, we consider and respond to a range of potential objections.

\section{The Colourful History of String Theory}

In the 1950s particle physics underwent a significant change; the development of large particle accelerators made it possible to create new hitherto unseen particles. These new particles posed peculiar novel problems for theorists: the particle types were too large in number and their properties (spin against mass squared) fell into patterns (such as the linear Regge trajectories represented on a Chew-Frautschi plot) that did not fit into any of the then standard frameworks provided by quantum field theory---indeed, many spins seemed too high to described by consistent quantum field theories. This led to the development of S-matrix based approaches, in which data coupled with axioms concerning the structure of the S-matrix were used to derive physical predictions. The final stage of this programme was the construction of the so-called dual resonance models, which were able to combine the various desirable properties of the S-matrix for strong interaction physics.\footnote{The model that achieved this is also known as the Veneziano model, since Gabriel Veneziano discovered it. The model is essentially just the Euler beta function: $A(s,t) = \frac{\Gamma(1 - \alpha(s)) \Gamma (1 - \alpha(t))}{\Gamma((1 - \alpha(s)) - \alpha(t))} =\int_{0}^{1} dx x^{-\alpha(s) - 1} (1 - x)^{-\alpha(t) - 1}$. The model violated unitarity in its original formulation, but this was later corrected.} The dual resonance model was soon seen to be derivable from a dynamical theory of strings.\footnote{The poles of the Veneziano amplitude (i.e. the singularities that occur when $s = m^{2} = t = \frac{n-1}{\alpha'}$) correspond to the string's mass spectrum. The interpretation provides an explanation of the infinite tower $J = \alpha M^{2}$ of mass-energy and spin states included in the Regge trajectories: they correspond to the oscillatory  (and rotary) modes of the strings.} However, it  had several features (the wrong particle spectrum, too many Lorentz dimensions, and other empirical inadequacies) that made it an unlikely candidate for describing the strong force. The emergence of `colour physics' proved to be the death knell of string theory as a theory of hadrons.

The two key vices of the early string theory, the particle spectrum problem (notably the existence of a massless spin-2 particle) and the expansion in the number of Lorentz dimensions (to cancel the conformal anomaly), were turned, by Jo\"{e}l Scherk and John Schwarz \cite{ss} (and independently, by Tamiaki Yoneya\footnote{In fact, though he was developing work instigated earlier by Jo\"{e}l Scherk (involving the $\alpha \rightarrow 0$ limit of dual models and the relation to Yang-Mills fields), Yoneya showed in 1973 \cite{yon,yon2}---a year earlier than the Schwarz and Scherk paper that is usually credited as the first `string theory as a quantum theory of gravity' paper---that certain dual resonance models (the Virasoro-Shapiro model) contained Einstein's theory of gravity as a zero slope limit. This paper marks the birth of string gravity, at least in the published record.}), into virtues of a new theory with a different target: (quantum) gravity. Though the new product emerged from a retuning of a free parameter (the string tension), we should really view this as an entirely new theory, for the intended systems that the theory is seen to apply to are entirely different in kind. Despite this, there are still general physical principles (originally seen to stem from physically necessary properties of interacting hadrons) that form the basic mathematical structure of the new theory.

The refashioning of the theory into a theory of gravity meshed well since both involve Riemannian geometry and both involve the imposition of dynamical conditions on the structure of spacetime. String theory is (or was) the theory of 2D Riemann surfaces (the worldsheets of the evolving strings). This brings with it copious amounts of extremely interesting mathematical machinery, having to do with conformal symmetry, modular invariance, algebraic geometry, and vertex algebras. The targeting of mathematical structures by physical principles and physical data (responsible for the overall structure of string theory) is not unique to string theory, of course. It is a general feature of physical theories that they pick out one or more mathematical structures that are used to represent physical systems described by the theory. The mathematical structures identified by string theory, however, are especially fruitful, as we will see.

\section{What is String Theory?}

At the outset we face a problem in defining string theory, since it comes in many forms. In this paper I assume we are talking about string theory as a unified theory of the four interactions of Nature, including a quantum theory of gravity. But even here we face a definitional problem: the theory has a known perturbative definition (where there are six such possible theories: one bosonic and five fermionic), but this is thought to be only part of the story. Though there is no complete non-perturbative definition, there are plenty of clues suggested by the various duality symmetries knitting the consistent perturbative expansions together. The five consistent superstring theories, for example, though apparently different at the perturbative level, look the same at tree level.

Given the incompleteness of the non-perturbative formulation, string theory is, then, usually presented in a perturbative fashion, expanding out string worldsheets in powers of the coupling constant $g_{s}$ of the theory.\footnote{I should perhaps point out that this perturbative `worldsheet' formulation is somewhat outmoded. However, it is at least well-defined and enables one to see in a fairly visual way how the interesting elements of mathematics (such as Riemann surfaces, modular invariance, and the like) enter into string theory and then find application in pure mathematics. Though I don't discuss it here, the modular invariance lies beneath some of the deepest connections between physics and mathematics, and is connected also to S-duality (the strong-weak coupling duality).} We restrict the discussion to closed string theory, and begin with the so-called $\sigma$-model. One wants to construct an action to describe the string dynamics in spacetime. The initial step is to consider a map $\Phi$ from a complex curve (a Riemann surface) $\Sigma$ representing the 2-dimensional string worldsheet\footnote{This worldsheet has a metric $h_{\alpha \beta}$ defined on it in the so-called Polyakov version. In the original Nambu-Goto version the worldsheet was metric-free. The surface also has a genus $g$ which plays a crucial role in the quantum theory.} into the ambient target space $X$ (with metric $G$ and additional background fields $B^{i}$):

\begin{equation}
\Phi : \Sigma \longrightarrow X
\end{equation}

\noindent The action is then a function of this map (including the worldsheet's metric), given the background fields $G$ and $B^{i}$:

\begin{equation}
S(\Phi , G , B^{i})
\end{equation}

\noindent The $\Phi$ field gives the dynamics of a 2-dimensional field theory of the worldsheet relative to the fixed background fields, one of which is the metric. The quantum theory (in 1st quantized form) is given by the path-integral (over moduli space: i.e. the space of inequivalent 2D Riemann surfaces, or Teichm\"{u}ller space):

\begin{equation}
\mathcal{P}(X) = \displaystyle\sum_{g} \int_{moduli_{g}} \int \mathcal{D} \Phi e^{i S(\Phi , G , B^{i})}
\end{equation}

\noindent In terms of the interpretation of this object, there is a degree of non-separability of the kind found in loop quantum gravity, for the relevant domain is not the space of metrics on a manifold (i.e. geometries) but the loop space. However, there are consistency conditions that must be met by string models not shared by the loop models. The most important of these concerns the restriction of the number of Lorentz dimensions in order to resolve the conformal anomaly: 26 (in the bosonic case) or 10 (in the bosons + fermions case). 

In order to preserve broad qualitative properties---such as the appearance of 4D spacetime, the empirical adequacy of general relativity at low energies---one needs to compactify on a manifold with a very stringent structure. Calabi-Yau manifolds are the candidates for the compact, internal manifold that are demanded by internal and external (i.e. phenomenological) consistency. It's invariant properties are responsible for determining the observable low energy physics in the non-compact, 4-dimensional manifold we ordinarily call spacetime. Let us spell out the details of this compactification strategy a little more, since it is utilised in the example of mirror symmetry that forms the basis of the central argument of this paper.

Quantum superstring theory remains Lorentz invariant only if spacetime has 10 dimensions. To construct a realistic theory therefore demands that the vacuum state (i.e. the vacuum solution of the classical string equations of motion, supplying the background for the superstrings) is given by a product space of the form $\mathcal{M} \times \mathcal{K}$, where $\mathcal{M}$ is a non-compact four dimensional Minkowskian spacetime and $\mathcal{K}$ is a compact 6-real dimensional manifold. One gets the physics `out' of this via topological invariants of $\mathcal{K}$ and gauge fields living on $\mathcal{K}$. One chooses the specific form of the compact manifold to match the observed phenomena in $\mathcal{M}$ as closely as possible.\footnote{The `Landscape Problem' is tantamount to the severe degeneracy in this (moduli) space of possible classical vacua.}

If one wants $\mathcal{N} = 1$ supersymmetry in the  non-compact dimensions $\mathcal{M}$, then one requires a very special geometry for the compact dimensions $\mathcal{K}$, namely a Calabi-Yau manifold mentioned above. This is defined to be a compact K\"{a}hler manifold with trivial first Chern class---this is just mathematical shorthand for saying that we want to get our low-energy physics (Ricci flatness\footnote{The first Chern class $c_{1}(\mathcal{X})$ of a metric-manifold is represented by the 2-form $1/2\pi \rho$ (with $\rho$ the Ricci tensor $R_{i\overline{j}} dz^{i} \wedge d\overline{z}^{\overline{j}}$). Calabi and Yau determined the various interrelations between Chern classes, K\"{a}hlericity, and Ricci forms. If one has a Ricci flat metric then one also gets the desired single supersymmetry since Ricci flatness is a sufficient condition for an $SU(3)$ holonomy group. Any textbook on complex algebraic geometry will explain these matters in detail---\cite{ballman} and \cite{kodiara} are good sources of information.} and the single supersymmetry) out of the compact dimensions.

There are five quantum-mechanically consistent superstring theories (in 10 dimensions: we ignore the purely bosonic case): Type I, $SO(32)$-Heterotic,  $E_{8} \times E_{8}$-Heterotic,  Type IIA and Type IIB. The Type I theory and the heterotic theories differ from the Type II theories in the number of supersymmetries, and therefore in the number of conserved charges. One is able to compute physical quantities from these theories using perturbation expansions in the string coupling constant. Given the extended nature of the strings, there is just a single Riemann surface for each order of the expansion (that is, the initially distinct diagrams can be topologically deformed into one another since there are no singularities representing interaction points: interactions are determined by global topological considerations of the world sheet (such as the number of handles), rather than local singularities).\footnote{Note, this is true for all but the Type I theory since its strings can be opened up. However, this does not need to concern us in what follows.} By looking at these expansions, in the `different' theories, one can find cases where the physics is identical so long as one makes transformations of a certain kind. Since these transformations are not taking us to a physically distinct state and relate states in different theories, they are referred to as `dualities'.\footnote{They have similarities with symmetries and gauge redundancies. However, with gauge redundancies we view the gauge related situations to represent one and the same physical state of affairs. In sting dual cases this does not seem possible since the `dual objects' can have distinct dimensionalities, sizes, and even differ as to whether they contain gravity or not, or are quantum mechanical or not.}

\section{T-Duality}

T-duality results from the combination of compactified dimensions and strings.\footnote{There are various options for the referent of the `T'. Some take it to refer to (T)arget space, the fact that it is similar to  the Kramers-Wannier (T)emperature duality of the Ising model, the fact that the theories that are T-dual are compactified on to (T)ori, or  to the fact that the letter `T' was used to refer to a low-energy field in early string theory.} T-duality is a kind of scale-invariance: it says that a theory at one size is equivalent to a theory at another size. It is essentially a duality that arises in conformal field theory. For superstring theories (i.e. with fermions and supersymmetry relating bosons and fermions) we find that the Type IIA and Type IIB theories are dual, as are the two heterotic theories. In the context of bosonic string theory it is a self-duality and can therefore be viewed as a gauge symmetry.

T-duality is very simply expressed: given two manifolds, with different compact geometries (in one of the spatial dimensions), a circle of $R$ and of radius $\tilde{R}$, and string length scale $\alpha'$, we have (schematically):

\begin{eqnarray*}
 \mathrm{String \, Theory \ on} \; R  \;  \;  \xleftrightarrow{isomorphic}  \; \;   \mathrm{String \, Theory \ on}  \;  \tilde{R} = \frac{\alpha'^{2}}{R}
\end{eqnarray*}

\noindent This isomorphism can be seen by considering the case where we have compactified one of the dimensions onto a circle. When this is done, the momentum is quantized around the circle according to the relation $p = n/R$ (where $n \in \mathbb{N}$). If we then consider the mass-energy of a system in such a compactified configuration then we must add a term corresponding to these so-called Kaluza-Klein modes:

\begin{equation}
E^{2} = M^{2} + \frac{n}{R}^{2}
\end{equation}

\noindent So far everything we have said applies just as well to particles. Strings have the additional property that they can wind around the compact dimension. This brings with it another term (the winding modes, where $m$ counts the number of such windings) that must be added to the total energy-mass:

\begin{equation}
\frac{1}{2\pi} \alpha' \times 2\pi R \cdot m = \bigg(\frac{mR}{\alpha'}\bigg)^{2}
\end{equation}

\noindent This gives us the following equation for computing the mass-energy:

\begin{equation}
E^{2} = M^{2} + \frac{n}{R}^{2} + \bigg(\frac{mR}{\alpha'}\bigg)^{2}
\end{equation}

\noindent If we then make the following (duality) transformations we leave the energy invariant:

\begin{eqnarray}
R \longrightarrow \frac{\alpha'}{R}\\
m \longleftrightarrow n
\end{eqnarray}

\noindent since we then have:

\begin{equation}
E^{2} = M^{2} + \frac{m}{\frac{\alpha'}{R}}^{2} + \bigg(\frac{n\frac{\alpha'}{R}}{\alpha'}\bigg)^{2}
\end{equation}

\noindent This can be converted back into the original by simply multiplying the numerator and denominator of the 2nd and 3rd terms by $R$ and cancelling the $\alpha'$s in the 3rd term.

Though this is a very elementary account, it serves to highlight the curious nature of strings and compact dimensions: \emph{from the stringy perspective there is no difference between a space with a large radius and one with a small radius!} If we consider a theory to be an equivalence  class of structures (with the equivalence given by the determination of identical observables) then what we took to be four distinct theories---type IIA and IIB on the one hand, and $SO(32)$ and $E_{8}\times E_{8}$ on the other---are really just two.

Physical sense can be made of this by viewing T-duality through the lens of the uncertainty principle: the attempt to localize a closed string at very small scales increases its energy-momentum. This increase in energy as one localizes to smaller and smaller length scales increases the size of the string.

In a nutshell, T-duality tells us that it is only some deeper intrinsic properties of the backgrounds for string propagation that matter in terms of `the physics'. Different background spaces are identical from the point of view of the strings. Since, in a string theory, everything is assumed to be made of strings, then in a purely string theoretic world, these backgrounds are indiscernible. This is very similar to the implications of diffeomorphism invariance in general relativity. There the \emph{localization} of the fundamental objects relative to the manifold  is a gauge freedom in the theory: the physics is therefore insensitive to matters of absolute localization. Quantities that are defined at points of the manifold are clearly not diffeomorphism-invariant, and therefore not gauge-invariant. The physics should not depend on such gauge-variant local properties. In the case of string theory, the physics should not depend on the size of the compact dimensions.

\section{Mirror Symmetry}

Mirror symmetry is one of the most conceptually curious aspects of string theory. It is essentially a generalization of T-duality (which holds only for homeomorphic manifolds) to topologically inequivalent manifolds. Recall that a phenomenologically respectable string theory requires that six of the 10 dimensions be hidden from view somehow. Compactification is the process that achieves this (at least formally). As we saw earlier, this  involves writing the 10 dimensional spacetime  $\mathcal{M}_{10}$ (required by quantum consistency) as a product space of the form $M^{4} \times K^{6}$, where $M^{4}$ is flat Minkowski spacetime and $K^{6}$ is some compact 6 real-dimensional space. $M^{4} \times K^{6}$ then forms the background space (the ground state in fact) for the classical string equations of motion. One chooses $K^{6}$ in such a way so as to use its geometrical and topological structure to determine the physics in the four non-compact spacetime dimensions (i.e. the low-energy physics). By choosing in the right way one can get explanations for a host of previously inexplicable features of low-energy physics, such as the numbers of generations of particles in the standard model, the various symmetry groups of the strong, electroweak, and gravitational forces, and the masses and lifetimes of various particles.

Calabi-Yau manifolds were found to be of importance in string theories since they allow for $\mathcal{N} = 1$ supersymmetries in four spacetime dimensions and other nice properties. Calabi-Yau manifolds are compact spaces satisfying the conditions of Ricci-flatness (to accommodate general relativity at the phenomenological 4D level) and K\"{a}hlericity (generating the $\mathcal{N} = 1$ supersymmetry in the non-compact dimensions). The problem is, there is a huge number of Calabi-Yau spaces (in D=6) meeting the required conditions, so the selection of one is a difficult task. However, what I want to discuss here is the identification of various of these, seemingly very different, manifolds via mirror symmetry.

To characterize manifolds one needs to know about their topological structure. To pick out this structure one looks for the invariants, of which there are various kinds. For example, a real 2-dimensional manifold is specified by its genus.  In string theory, the topological and complex structure of the compact manifold determines the low energy physics in the real, four non-compact dimensions. What was required by the string theorists, in order to consistent the observed  particle physics, was a Calabi-Yau space with an Euler characteristic $\chi$ of $\pm 6$. These can be found (and were found by Yau himself). However, there is an entire family of `mirror' Calabi-Yau spaces with opposite Euler number. These look distinct from a topological and complex structure perspective, but from the point of view of the string theory (or, more precisely, the 2D conformal field theory) living on these spaces, the difference is merely apparent: the field theory is insensitive to the mirror mapping and is, in this sense, background independent.

\begin{figure}
  \begin{center}
   \includegraphics*[width=10cm]{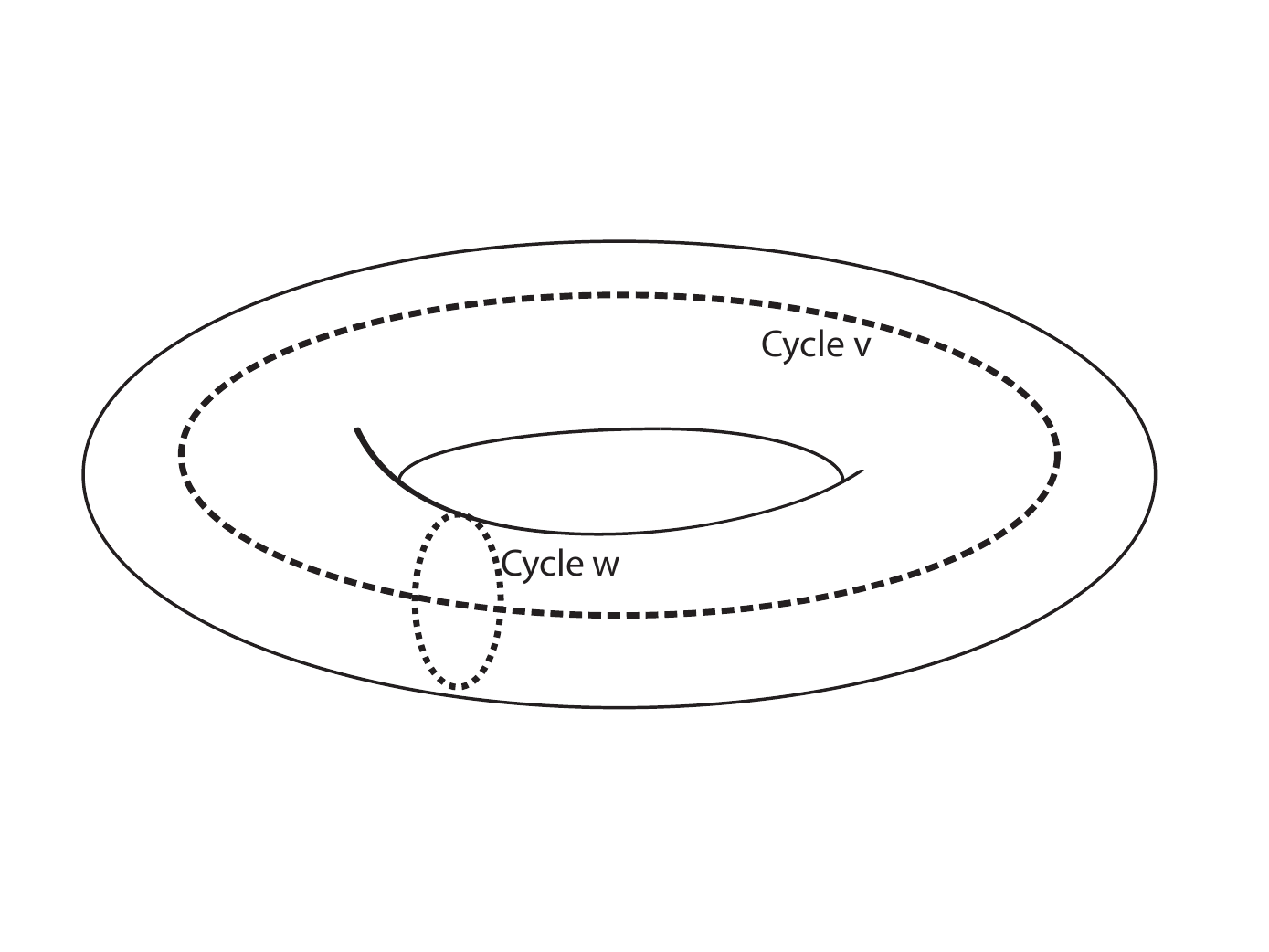} \end{center} \caption{The two independent cycles of a torus.} \label{cycles} \end{figure}

The concept of the Hodge diamond makes the phenomenon of mirror symmetry easy to see in a visual way, and was in fact discovered and named as a result of this visual appearance.  Hodge numbers are to (complex) K\"{a}hler manifolds what Betti numbers are to real manifolds: they specify topological invariants of the manifold and correspond to the dimension of the relevant cohomology group. The Betti numbers count the number of irreducible $n$-cycles of some manifold---see fig.\ref{cycles}. 

The $n$-cycles themselves  are defined as `chains' without boundary, where chains are sums of (oriented) submanifolds of the manifold. So, for example, $H_{n=0}$, a 0-cycle is a 0-chain and and is simply a point---note, cycles are considered equivalent if they differ by a boundary; so, for example, for a \emph{connected} manifold, all points are deemed equivalent. The Hodge numbers do the same, but for complex cycles $p$ and their complex conjugates $\overline{p} = q$.
Schematically: 

\begin{quotation}
$\mathrm{DeRham \, Cohomology \, Group} \,  H^{n}_{D}  \Rightarrow \mathrm{Betti \,number} \; b_{n} = \mathrm{dim}(H^{n}_{D})\\
\mathrm{Dolbeault \, Cohomology \, Group} \,  H^{p,q} \Rightarrow \mathrm{Hodge \,number} \;  h^{p,q} = \mathrm{dim}(H^{p,q})$
\end{quotation}

\noindent The Betti number and the Hodge number are related (by the Hodge decomposition) as:

\begin{equation}
b_{n} = \displaystyle\sum_{p+q = n} h^{p,q}
\end{equation}

\noindent The Hodge diamond encodes these various Hodge numbers as follows:

\begin{eqnarray}
\begin{array}{ccccccc}
 & & &h^{0,0} && &\\
&& h^{1,0} && h^{0,1} && \\
& h^{2,0} && h^{1,1} && h^{0,2} & \\
h^{3,0} && h^{1,2} && h^{2,1} && h^{0,3} \\
& h^{1,3} && h^{2,2} && h^{3,1}& \\
&& h^{2,3} && h^{3,2} && \\
& &&  h^{3,3} & &&
\end{array}
\end{eqnarray}

\noindent For a complex 3-dimensional manifold, we can compute the Hodge numbers via the Hodge decomposition, giving:

\begin{eqnarray}
\begin{array}{ccccccc}
b_{0} =1  \\
b_{1} = 0 \\
b_{2} = h^{1,1} \\
b_{3} = 2(1+h^{2,1}) \\
b_{4}= h^{2,2} = h^{1,1} \\
b_{5}= 0 \\
b_{6}= 1 \\
\end{array}
\end{eqnarray}

\noindent The only independent Hodge numbers of the 3-manifold (with non-vanishing Euler characteristic---see below) are $h^{1,1}$ (roughly describing, via a number of real parameters, the size, or radius, and shape of the manifold) and $h^{2,1}$ (roughly the number of complex parameters to describe the complex structures that can be defined on the manifold). The other numbers are set by various mathematical identities and properties: $h^{p,q} = h^{q,p}$ by complex conjugation; $h^{p,q} = h^{3-p, 3-q}$ by Poincar\'{e} duality (giving us the identity $h^{1,1} = h^{2,2}$ above); and the condition of vanishing first Chern class sets up an isomorphism between $h^{0,p}$ and $h^{0, 3-p}$. Hence, we have:

\begin{eqnarray}
\begin{array}{ccccccc}
 & & &1 && &\\
&& 0 && 0 && \\
& 0 && h^{1,1} && 0 & \\
1 && h^{2,1} && h^{2,1} && 1 \\
& 0 && h^{1,1} && 0& \\
&& 0 && 0 && \\
& &&  1 & &&
\end{array}
\end{eqnarray}

\noindent Since the Euler number $\chi$ for a real manifold is computed via the Betti numbers as:

\begin{equation}
\chi = \displaystyle\sum_{n} (-1)^{n} b_{n}
\end{equation}

\noindent The Euler characteristic for a complex K\"{a}hler manifold can be computed, again invoking Hodge decomposition, as:

\begin{equation}
\chi_{k} = \displaystyle\sum_{p,q} (-1)^{p+q} h^{p,q}.
\end{equation}

\noindent This number is, as mentioned above, crucial in the mapping to real-world, low-energy physics.

It is a claim of algebraic geometry, having its origin in string theory, that every space described by such a Hodge diamond has a mirror (with the axis of reflection lying along the diagonal). The phenomenon of mirror symmetry then refers to an isomorphism between pairs of conformal field theories (worldsheet string theories) defined on prima facie very distinct Calabi-Yau manifolds, differing even with respect to their topology. In this case the manifolds have their Hodge numbers switched as:

\begin{equation}
H^{p,q} (M) \xleftrightarrow{isomorphic} H^{n-p,q} (\tilde{M})
\end{equation}

\begin{figure}
  \begin{center}
   \includegraphics*[width=10cm]{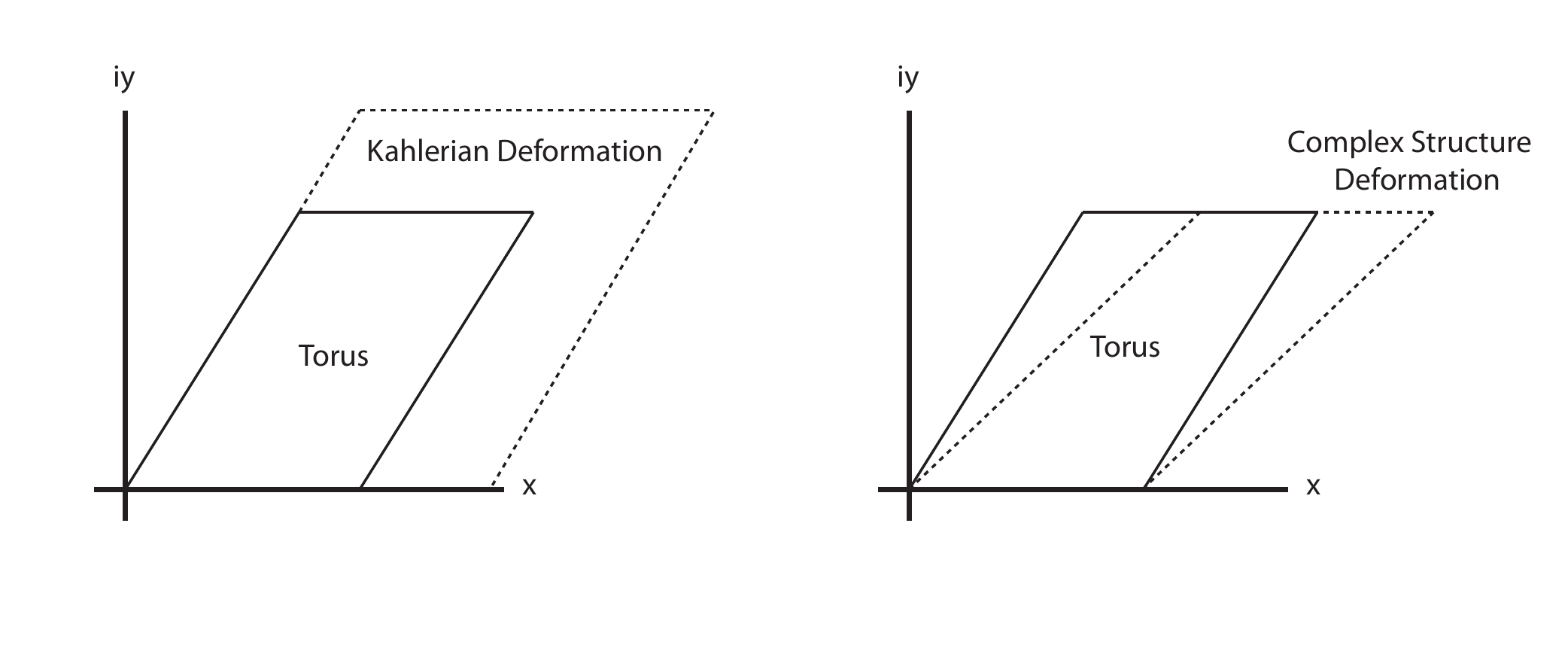} \end{center} \caption{The torus (with top and bottom and left and right identified) is an example of a 1-dimensional Calabi-Yau manifold. Deformations of the K\"{a}hler form of the torus change the volume while leaving the shape invariant (that is, the angles between the independent cycles are constant). A complex structure deformation does the opposite: it changes the shape (the angles) while leaving the volume invariant. (Adapted from Greene \cite{greene}, p. 25.)} \label{tor} \end{figure}

\noindent Where $n$ is the (complex) dimension of the manifold. In the case where this is 3, we find that the remaining Hodge numbers $h^{1,1}$ and $h^{2,1}$ are isomorphic. These numbers parametrize  the size and shape of the compact space, along with its complex structural properties---see fig.\ref{tor}.\footnote{They correspond to topologically nontrivial 2-cycles and 3-cycles respectively.} Mirror symmetry tells us that the physics (of relativistic \emph{quantum} strings) is invariant when these, apparently very different (with different corresponding \emph{classical} theories), features are exchanged. That is, there is quantum equivalence despite a marked difference at the classical level.

For example, the Euler character is equal to twice the number of particle generations. It can be connected to these shape and size parameters as follows:

\begin{equation}
\frac{|\chi|}{2} =  |(h^{1,1} + h^{2,1})| =   |(h^{1,1} - h^{2,1})| = \frac{|-\chi|}{2} = \mathrm{No. \,Gen.}
\end{equation}

\noindent To achieve a realistic string theory, then, one needs to find a Calabi-Yau manifold with $h^{1,1}$ and $h^{2,1}$ satisfying:

\begin{equation}
|(h^{1,1} + h^{2,1})| = 3
\end{equation}

\noindent Gang Tian and Shing-Tung Yau discovered such a manifold \cite{yau}. Though there is degeneracy here too, with multiple candidates available.

\section{Using Mirrors to Count Curves}

This setup was used to great (and surprising) effect to resolve a problem in pure mathematics, in the field of enumerative geometry. Briefly,  what are now known as Gromov-Witten invariants (interpreted as topological string amplitudes) were used to calculate the number of curves of a given degree of a particular surface.\footnote{Full and very readable accounts of mirror symmetry, including the application discussed in the subsection, can be found in: \cite{mir1} and \cite{mir2}. An excellent, elementary overview, including aspects of enumerative geometry, is \cite{katz}.} Using string theory, Candelas \emph{et al.} \cite{candy} developed a generating function to find the number of curves $n$ for all degrees $d$ through a particular surface, a well-known Calabi-Yau complex 3-manifold known as a quintic (in fact, the simplest possible Calabi-Yau manifold), defined by the equation: 

\begin{equation}
x_{o}^{5} + x_{1}^{5} + x_{2}^{5} + x_{3}^{5} + x_{4}^{5} = 0 \supset \mathbb{P}^{4}
\end{equation}

\noindent The function they came up with was based on string perturbation theory (that is, a sum-over-Riemann-surfaces approach):

\begin{equation}\label{n}
K(q) = 5 + \sum_{d=1}^{\infty} n_{d} d^{3} \frac{q^{d}}{1-q^{d}}
\end{equation}

\noindent Mathematically, $n_{d}$ is the number of rational curves of degree $d$, and $q = e^{2\pi i t}$. In terms of the physics, $n_{d}$ is the `instanton number', pertaining to the quantum corrections.\footnote{In more rigorous accounts, $n_{d}$ is taken to represent the Gromov-Witten invariants of the space. These, roughly, correspond to the structure that is left invariant under deformations of the complex structure (i.e. those infinitesimal deformations parametrized by the cohomology group $H^{2,1}$).} Each curve of degree $d$ adds $d^{3} \frac{q^{d}}{1-q^{d}}$ to the Yukawa coupling.  This spits out the various intersection numbers  as coefficients in the expansion:

\begin{equation}
K(q) = 5 + 2875 \frac{q}{1-q} + 609250 \cdot 2^{3} \frac{q^{2}}{1-q^{2}} + 317206375 \cdot 3^{3} \frac{q^{3}}{1-q^{3}} + \cdots
\end{equation}

\noindent That is

\begin{eqnarray*}
n_{1} = 2875\\
n_{2} = 609250\\
n_{3} = 317206375\\
\vdots
\end{eqnarray*}

\noindent The $d = 1$ and $d = 2$ cases were already well-known. But $d = 3$ was under investigation. The string theoretic calculation turned out to be correct, giving strong evidence that the formula was giving the correct values for other values.\footnote{I discuss the methodological ramifications of this scenario (\emph{vis-\`{a}-vis} the concept of \emph{evidence} for string theory) in \cite{ricklesmir}. Peter Galison has a related, though more historical article, covering similar themes: \cite{galison}.}

The application of the duality (mirror symmetry) here amounts to the `simulation' of the difficult \emph{quantum} corrections (which yield the desired intersection numbers as instanton corrections) using aspects of the \emph{classical} geometry in the dual theory. As Vafa explains:

\begin{quotation}
\noindent [W]hat happens is that a parameter which controls quantum corrections $\lambda_{0}$ on one side gets transformed to a parameter $\tilde{\lambda}_{k}$ with $k \neq 0$ describing some classical aspect of the dual side. This in particular implies that quantum corrections on one side have the interpretation on the dual side as to how correlations vary with some classical concept such as geometry.
(\cite{vafa}, p. 540)
\end{quotation}

\noindent In the case of the string theoretic enumerative geometry, what is going on is that the Yukawa coupling (here, the 3-point vertex function or correlation function) is giving the count of the curves. This function contains both a classical (easy) piece and a quantum corrected (hard) piece. Following the prescription sketched by Vafa above, one can compute the quantum part using elements of the classical geometry and then convert back.  Given the remarkable nature of this application of the duality conjecture, one might not unreasonably view the positive results as offering evidence for the correctness of the duality and the theory to which it belongs, namely superstring theory.\footnote{The formula of Candelas \emph{et al.} was in fact made more rigorous using a variety of techniques external to string theory, see \cite{lian} for example. The various \emph{proofs} of mirror symmetry can be found in \cite{mir2}. The establishment of the mirror symmetric counting of curves depended precisely on the formalization of the instanton corrections using the tools of Gromov-Witten theory.} I shall attempt to defend such a view (making an inference from novel mathematical predictions to physical theory) in the following two sections.

\section{Mathematical Miracles and Scientific Methodology}\label{meth}

In his \emph{The Trouble with Physics}, Lee Smolin writes that ``[d]espite the absence of experimental support and precise formulation, the theory is believed by some of its adherents with a certainty that seems emotional rather than rational'' (\cite{smolin}, p. xx). Smolin is not convinced that string theory's ability to generate mathematical results is enough to justify its level of support within the physics community. For Smolin, what is needed are tests: concrete, physical tests---or, at least the ability to suggest potential tests. Is it true that string theorists' often strong adherence to their theory is \emph{irrational}? I argue not.

\subsection{From Cosmic Coincidence to Realism}

In the context of realism about scientific theories, J. J. C. Smart argued that instrumentalists (i.e. anti-realists) about scientific theories must believe in ``cosmic coincidence''. He puts the point this way:

\begin{quotation}
\noindent Is it not odd that the phenomena of the world should be such as to make a purely instrumental theory true? On the other hand, if we interpret a theory in the realist way, then we have no need for such a cosmic coincidence: it is not surprising that galvanometers and cloud chambers behave in the sort of way they do, for it there really are electrons, etc. this is just what we should expect. A lot of surprising facts no longer seem surprising. (\cite{smart}, p. 39)
\end{quotation}

\noindent Belief in the \emph{truth} (or approximate truth) of an empirical successful scientific theory is, on this argument, the only stance that does not make that success puzzling. Hilary Putnam \cite{putnam} labels this the ``no-miracles argument''.

The argument begins with some puzzling fact (some phenomenon that cannot otherwise be accounted for). It is then noted that this fact can be derived as a theorem of some theory, and this is finally taken as \emph{evidence} for the theory that generated that fact.\footnote{Bas van Fraassen \cite{bas} has provided a convincing `Darwinian' anti-realist counter-argument to this no-miracles argument: ``any scientific theory is born into a life of fierce competition, a jungle red in tooth and claw. Only the successful theories survive---the ones which \emph{in fact} latched on to actual regularities in nature'' (p. 39). Success here is just a selection-effect, analogous to the fitness of an organism: theories not well adapted to their environment (i.e. the actual regularities in nature) are quite naturally rejected. Truth \emph{per se} is playing no role in success.} What I am suggesting in this paper, though disanalogous in many ways, is that there is something like this `inference to the best explanation'-style argument supporting belief in string theory. 

But clearly (and this is the most glaring disanalogy) success in the string theoretic context cannot be \emph{empirical} success in the sense of accurate physical predictions about real-world string-theoretic systems. Here I am understanding success to be mathematical success. That something like this thought grounds the adherence of string theorists to their programme can be seen, for example, in a recent book on string theory (and other themes), in which David Olive writes that ``the physical ideas [in string theory] have gained support from the startling and successful repercussions they have had in pure mathematics in terms of conceptual breakthroughs'' (\cite{ol}, p. 3). This looks like a case of mathematical support of a physical theory. There are ways of making a case for this, on the surface, unlikely argument. I begin with two potential objections.

\subsection{The Mathematical Fertility of False Theories}

Lee Smolin (\cite{smolin}, pp. 34--35) discusses the origins of knot theory in Thomson's (i.e. Lord Kelvin's) study of classical electromagnetism \cite{thom}, developing from the idea that atoms were knots in magnetic field lines (making them vortex tubes of ether). It was actually P. G. Tait who abstracted from this the mathematical theory of knots, amounting to an in depth classification of possible knots.   

Smolin argues that if mathematical fertility could be an indicator of truth, then we ought to take the success of knot theory as evidence for the idea that atoms are indeed knotted bits of ether. Hence, we have an apparent \emph{reductio ad absurdum} of the idea that I am arguing for in this paper, that mathematical fertility---such as the postulation of mirror symmetry and the resolution of difficult problems in enumerative geometry---might lead us to believe more strongly in a theory.

There are several points we can make in response. Firstly, I am not arguing that mathematical fertility weighs as heavily as a good empirical confirmation (or empirical disconfirmation) in our evaluations of physical theories. Secondly, that Kelvin's theory was eventually disconfirmed does not mean that it was a \emph{bad} theory---after all, it was discussed and studied as a serious theory for some 20 years. It was precisely the fact that it was taken seriously as a physical theory that led to the development of knot theory. That it was taken seriously for 20 years means that, structurally at least, the theory was `on to something': it got something right, just as Newton's theory of gravitation, though incorrect at certain scales and energies, still gets something right. Thirdly, and related to the second point, the physics of knots did not vanish forever after this failed episode. Rather, it forms an integral part of modern physics, especially in condensed matter physics, quantum field theory, and quantum gravity (Smolin, more than most, knows this well, of course). 

Hence, what Kelvin latched on to was some widely applicable piece of representational machinery. I think we \emph{can} say that the development of knot theory \emph{ought} to have given reason to believe in Kelvin's theory \emph{in the absence of a better confirmed theory or weightier disconfirming evidence}.

\subsection{The Causal Isolation of Mathematics}

The argument from mathematical predictions appears to fall foul of the `concrete versus abstract' division separating physics from mathematics. Mathematics is causally isolated: it's objects are non-spatiotemporal. So how can it be that mathematical results can have any impact on physical facts? This immediately presupposes that Platonism is involved. This is not a necessary consequence. We could, conceivably, adopt a conventionalist or some other viewpoint. J. S. Mill, for example, espoused a curious empiricism about mathematical truths, viewing them as extremely general laws of nature. For example, the laws of addition will be satisfied by physical objects. Frege argued that this position conflated mathematics with its application. However, Lakatos later resuscitated Mill's idea. He argued that mathematical theories, just like physical theories, were fallible.

Elliot Sober \cite{sober} has argued that there could never be the kind of relationship between physical and mathematical that I have been arguing for on the grounds that, while we would be willing to drop a claim about the physical world on the basis of empirical evidence, we would not do the same with mathematical claims. Mathematical truth is, as Mark Colyvan puts it, ``never placed on the line'' (\cite{coly}, p. 114). Mathematics does not simultaneously get `tested' by experiments that test some physical hypothesis. But if this is the case, then if mathematics gets to join in the success of physical theories, then it should also suffer the failures. Of course if a theory fails we don't assume the mathematical theory used to describe it was wrong, but simply that it was wrongly applied. 

There is a response we can make here: provided that the physical theory is consistent, we can say the same about it. That is, if only the universe were put together in the right manner, the theory would be perfectly applicable. Note also that there has recently been a spate of arguments in the philosophy of mathematics literature arguing that there can be purely mathematical explanations of physical phenomena---see e.g. \cite{baker}. If these arguments are correct then there can be crossings here too. However, this argument has come under fire precisely on the grounds that it doesn't close the causal gap from mathematical to physical facts. Sorin Bangu (\cite{bangu}, p. 19).

Is there a way even for those who espouse predictivism (that a theory gains more support from evidence that it was not designed to fit) to grant sound scientific status to string theory, and perhaps rank it more highly than other approaches to quantum gravity? I believe I have given one example already: the case of mathematical predictions that were not expected and that did not enter into the construction of the theory.  It is true that this example, involving mirror symmetry, is not the kind of phenomenon that could be tested in the laboratory. However, such predictions can be tested and have been tested (and found to be correct) using computer simulations. In other words, string theory leads to  mathematical predictions (about what are essentially represented spacetime structures, and the dynamics of strings in such spaces) that are testable using computers. In high energy contexts, or impractical situations, this is a perfectly legitimate methodology. Much of what we know of QCD, black hole physics, and indeed general relativity, is exactly of this kind, and we don't think any less of general relativity or QCD for the fact that this is so.

\subsection{The Social Isolation of Mathematics}

An alternative line of objection comes from Penelope Maddy, who argues that if there really were the kind of interaction between mathematics and physics that I have been proposing, then we ought to see mathematicians taking a vital interest in physics:

\begin{quotation}
\noindent If this were correct, one would expect set theorists to be vitally interested in the implications of renormalization in quantum field theories, in developments in quantum gravity, in assessments of the literalness of other applications of continuum mathematics in natural science, for the propriety of their very methods would hang in the balance. (\cite{pen}, p. 159)
\end{quotation}

\noindent But, she notes, set theorists couldn't care a less: they are socially isolated from physics. Firstly, this might be true of some mathematicians, but it certainly isn't true for all, or (I expect) even a large portion. I can think of mathematicians with interests in physics who are interested in the physical ramifications for category theory, for example. Geometry received an enormous impulse from the interactions with physics. 

Secondly, she argues that the supposed indifference comes from the fact that there would be a \emph{practical} indifference in the way set theorists would work: the methods would be unchanged. But again, this doesn't strike me as correct. Physical applications often have back-reaction on mathematics.  Geometry again offers a counterexample. Yang-Mills theory has provided many new tools for mathematicians. Moreover, what the mirror symmetry example has shown is that if methods in the natural sciences are able to deliver \emph{results}, then mathematicians will take note, and it could well infect their methods. Maddy's argument rests, then, on too restricted a view of mathematics and mathematical physics.

\subsection{Quinean Holism and the Indispensability Argument}

A stronger link between the mathematical and the physical comes from the so-called indispensability argument originating with Willard Quine. Quine argued for holism about knowledge: belief in some hypothesis or theory is justified if the hypothesis or theory forms part of, or coheres with our overall knowledge. His own capsule formulation of this idea is expressed as follows: ``our statements about the external world face the tribunal of sense experience not individually but only as a corporate body'' (\cite{quine}, p. 41). Furthermore, since our best physical theories are dependent on the truth of the mathematics than one uses to formulate them, empirical confirmation of the physical theory is just as much an instance of empirical confirmation of the mathematics. Mathematical objects are theoretical entities, just like electrons and quarks!\footnote{I am grateful to David Armstrong for bringing the potential relevance of Quine's position to my attention.}

This has direct consequences for the `abstract' versus `concrete' division: physical experiments that successfully confirm some theoretical prediction, where that theory is (indispensably) linked to some piece of representational machinery, likewise confirm the mathematics. If we think once again in terms of the no miracles argument, then if the successful prediction gives us some reason to believe in the existence of the entities that the theory uses to make the prediction, then (by parity of reasoning) it also gives us reason to believe in the existence of mathematical entities.  
This holistic view of confirmation quickly leads, then, to what is known as the ``indispensability argument'': mathematics used in the construction of a theory receives empirical support, just as much as the theoretical entities used.\footnote{Naturally the negative responses to this thesis have tended to argue that mathematics is not indispensable to science, and our usage of it in this context is nothing but a matter of convenience (see, e.g. \cite{field}).}

But now what is stop us inverting this argument, and arguing instead that mathematical confirmations (derived from some `physical' theory) can act as confirmations of the physical theory? In this inverted Quinean argument, then, we use successful mathematical predictions as support for a physical theory. There is no asymmetry in the direction of support; at least it is hard to see what could account for such an asymmetry if we are willing, as Quine does, to allow empirical confirmations of mathematical truths. If we can speak of physical evidence for mathematical objects and truths, then there ought to be room for mathematical evidence of physical facts. There is an obvious sense in which this is perfectly true: blatant mathematical inconsistency will enable us to infer that the physical world will not be able to instantiate it.

This possibility is made somewhat more palatable by the fact that the mathematical structures in question, in the case of string theory, are isolated by physical principles. String theory is, as Polchinski puts it, ``a mathematical structure...deeply grounded in physics'' (\cite{polch}, p. 429). The fact that this structure is able to generate so many mathematical discoveries tells us something about the physical theory too. And as Putnam puts it: ``if we were really just writing down strings of symbols at random, or even by trial and error, what are the chances that our theory would be consistent, let alone mathematically fertile?'' (\cite{putnam}, p. 73).

\section{Accommodation versus Prediction}

Let us now turn back to the issue of testability that has been laying fallow until now, for those who defend something like the view defended by Smolin may still have doubts about the rationality of adopting a physical theory that has made no directly  testable predictions. The crux of the issue for string theorists is the debate between accommodationism (or, somewhat stronger, `explanationism') and predictivism. The traditional view is that the ability of a theory to make predictions of novel phenomena (not used to guide the construction of the theory) weighs more heavily than its ability to explain old phenomena. However, the historian of physics Stephen Brush has marshalled several case studies that show this not always to be the case in practice. There are episodes in which the explanation of previously known \emph{but puzzling} phenomena weighed more heavily than novelty of predictions.\footnote{Above I argued that some of the `unforeseen predictions' can be essentially mathematical, rather than physical. However, I will put this aside for the purposes of the discussion in this section.}

As Peter Achinstein \cite{ach} notes, this brings in some interesting \emph{historical} elements into theory evaluation: notions of degree of support, and related notions, are not time-independent. There are several possibilities. For example, given evidence $E$ and theory $T$:

\begin{itemize}
\item $E$ offers evidential support for $T$ iff $E$ was not known before $T$ \cite{mus}
\item $E$ offers evidential support for $T$ iff $T$ was not devised to explain $E$ \cite{zah}
\item $E$ offers evidential support for $T$ iff $E$ was not explainable by other theories before $T$ \cite{mus}
\end{itemize}

\noindent I don't wish to get involved in the difficult debate over whether evidence is or should be seen as historical or not. It is clearly true that those who take issue with string theory's claim to `retrodict' certain facts adopt a historicist  position.

In a series of articles \cite{brush1,brush2,brush3}, Brush  defends the view that novelty in predictions \emph{as a matter of historical fact} do not play a greater role in theory evaluation than explanation and accommodation of old, yet puzzling data. His central case study throughout has been the acceptance of general relativity. The standard story tells how scientists and the public were instantly converted by the confirmation of general relativity's light bending prediction. However, as Brush argues, general relativity was widely accepted \emph{before} this test, and was done so on the basis of its ability to get the perihelion of Mercury correct. This was so, argues Brush, despite the fact that this data was guiding the very construction of Einstein's theory, acting as a phenomenological target.  General relativity's retrodiction of Mercury's (up until then, anomalous) perihelion, though not a novel prediction, was novel in the sense that it was the only theory able to do so. In this case, then, it seems that \emph{uniqueness} of retrodiction is playing a crucial role.

Turing now to string theory. Many \emph{known} aspects of particle physics are inexplicable using currently established theories (or unestablished theories for that matter). For example: why are there three families of particles? Why are the particle's interactions governed by these particular symmetry groups? Why do we find the symmetries broken at these particular scales? In the context of string theory, these are delivered through the topology of $\mathcal{K}_{6}$. As are (via the Yukawa couplings), the particle lifetimes and masses. Though certainly not perfect, string theory does deliver a landscape (i.e. an ensemble of theories) with regions that correspond to something like the standard model. The topological features of the worldsheets combined with those of the background enable the generation of low energy physics with the right gauge groups, and the right number of generations, and more). As Schellekens notes (\cite{schell}, p. 11), though this is often seen as a trivial victory (given the vast size of the landscape) it isn't all that trivial since infinitely many other gauge theories are simply ruled at as physically impossibilities. So string theory has a case that though it is only an instance of accommodation rather than prediction, it is at least a case of unique accommodation.

Peter Lipton (\cite{lipton}, p. 21) has objected to Brush's account, noting how Halley's theory of comets made three distinct accommodations of known cometary trajectories before the prediction of the returning comet that bears his name was confirmed. A single novel prediction massively outweighed the three accommodations. We can ask two questions about this: (1) is this generic? (2) is it rational?  Lipton does find another case, involving the prediction of as of then unknown elements using the period table. He argues, again, that these were worth more than the accommodation of all the other elements. Stephen Brush (\cite{brush2}, p. 139) has taken Lipton to task for the lack of historical evidence for his claim. The burden of proof is on the predictivist to demonstrate that the confirmation of the prediction swayed scientists' opinions. But, as with general relativity, the theory was already accepted by the time the novel  prediction was confirmed. This is not to say that novel predictions  could never and do never play a crucial role in the evaluation of theories. But it does show that the story is not simple. A cursory inspection of the history of quantum gravity research shows quite clearly that novel predictions are not always involved in the acceptance and rejection of theories.

I mentioned that the uniqueness of retrodiction or accommodated facts might also be playing a vital role in the credence assigned to string theory. Richard Dawid has argued for string theory on the basis that it is unique \emph{simpliciter} \cite{daw}. It is, he argues, the only possible theory that does the job of unifying the forces. Let us suppose, for the sake of argument, that this was indeed true, that string theory is the \emph{only possible way} to bring together the forces of nature. Would this, in itself, make the theory `true'? If we are certain that there are just these four forces then it looks like it might have to be the case. But there is no definitive reason, still, that the forces must be unified. If there were, we could rule out any program in quantum gravity that seems only to quantize gravity (i.e. independently from the other interactions). Furthermore, I see no reason why even a theory of everything (for which, \emph{per impossibile}, we are certain that it gives a complete description of reality) we must suppose that it must be unique. There is no reason why there could not be multiple distinct frameworks for describing the same picture, even when we are dealing with `theories of everything'.

Aside from this, the central problem with this suggestion (that uniqueness can be an indicator of truth) is that it amounts to a claim without support from the theory---\emph{cf.} \cite{hed}, \S4, for a more general discussion of the problems with the uniqueness argument. String theory originated from Geoff Chew's bootstrap approach, and it was thought to provide, in its early stages, a unique bootstrap. This was responsible for much of the excitement that gathered around the theory. However, the uniqueness quickly degenerated in several ways. Firstly in the several different types of string theory, and then in the number of possible ways of compactifying them.  It is often said by string theorists that uniqueness is achieved by the duality symmetries that connect these theories, or that the various theories are ground states of one and the same theory, but this is wishful thinking: there is no internal reason to adopt this viewpoint. Indeed, duality symmetries are generally taken to relate \emph{distinct} theories, making them distinct from gauge redundancies.

\section{String Theory's Conscilience of Evidence}

Taken individually, string theory's instances of confirmation are admittedly relatively weak. We can enumerate at least five distinct categories\footnote{One might follow Dirac \cite{dirac} and add `beauty' to this list of evaluative factors. See \cite{mcal} for a thorough analysis (and defence) of the role of aesthetics in the evaluation of physical theories. For an analysis of the problems with taking beauty seriously in this way, see \cite{engler}.}:

\begin{enumerate}
\item Unification (`accomodation')
\item Universal structure
\item Simulations
\item QG Targets
\item Fertility
\end{enumerate}

\noindent Since string theory scores highly when we combine these diverse categories, then it scores highly overall \emph{given the absence of a competing theory that has made a well-confirmed experimental prediction}. William Whewell gave such numerosity of evidence (in the sense of how much a hypothesis explains) a central role in his approach to theory evaluation, labelling the feature ``consilience'' (\cite{whew}, p. 65). One can use the notion to compare competing theories, even in cases where there is no experimental evidence. One chooses the `more consilient' theory---though of course, one would have to factor in some kind of quality control on the kinds of facts that are explained, to rule out trivialities and such like.

Since we have, in previous episodes in science, always had the availability of experimental tests we have never really had to weigh these alternative theoretical virtues. However, we can find some such instances. General relativity had no competitors when it came into being. Those other theories of gravitation that existed were know to be empirically inadequate. In this case, even before the classic tests of the theory it was considered to be well-confirmed on the basis of old evidence.

Interestingly, Charles Darwin argued for the theory of evolution by natural selection using such a consilience of evidence, and faced much the same objections as string theorists face today. Darwin staunchly defend the method, writing in the 6th edition of \emph{The Origin of Species}:

\begin{quotation}
\noindent It can hardly be supposed that a false theory would explain, in so satisfactory a manner as does the theory of natural selection, the several large classes of facts above specified. It has recently been objected that this is an unsafe method of arguing; but it is a method used in judging the common events of life, and has often been used by the greatest natural philosophers. (\cite{darwin}, p. 476)
\end{quotation}

\noindent More interestingly, Karl Popper was one of those who would dispute the scientific status of Darwin's theory. He viewed it as a metaphysical system (though not an unworthy one). 

In a recent appraisal of string theory, Nancy Cartwright and Roman Frigg draw attention to the range of factors other than testability that can play role in our evaluation of theories. They also explain how string theory does well in some of these other ``dimensions''. But they still come down negatively on the status of string theory, arguing that:

\begin{quotation}
\noindent a research programme that progresses only in some dimensions, while being by and large stagnant in the others, surely does not count as being progressive. Contrasting string theory with Maxwell's unification of electricity and magnetism, for example, we can see that the latter was genuinely progressing and eventually successful in every dimension. It used the new and powerful concept of a field, which made the theory simple and elegant, while at the same time giving rise to a whole set of new phenomena that led to new predictions. (\cite{carf}, p. 15)
\end{quotation}

\noindent The conclusion Cartwright and Frigg draw from their analysis strikes me as a \emph{non sequiter}. After pointing out several ways in which string theory is progressive, they claim that nonetheless the theory is in fact degenerative or stagnant! Of course, we have something very different with string theory, and to compare it to Maxwell's theory, which made predictions well within energy capabilities of the day, is not helpful. 

In the final analysis, Cartwright and Frigg defend, more or less, the traditional view of scientific method:

\begin{quotation}
\noindent The question of how progressive string theory is then becomes one of truth, and this brings us back to predictions. The more numerous, varied, precise and novel a theory's successful predictions are, the more confidence we can have that the theory is true, or at least approximately true (see box). That a theory describes the world correctly wherever we have checked provides good reason to expect that it will describe the world correctly where we have not checked. String theory's failure to make testable predictions therefore leaves us with little reason to believe that it gives us a true picture. (ibid.)
\end{quotation}

\noindent As I mentioned at the outset of this paper, string theory (and quantum gravity research in general) simply cannot be bound to these same constraints. Inasmuch as it can, it is along much more indirect channels, such as it's performance in  simulations, it's ability to be applicable beyond its intended domain of application, and its history of generating mathematical results. But this is not sufficient for Cartwright and Frigg:

\begin{quotation}
\noindent Although string theory has progressed along the dimensions of unifying and explanatory power, this in itself is not sufficient to believe that it gives us a true picture of the world. Hence, as it stands, string theory is not yet progressive because it has made progress only along a few of the many dimensions that matter to a research programme's success. (ibid.)
\end{quotation}

\noindent The problem this passage exposes here is that Cartwright and Frigg slide from the evaluation of theories (not whether they are necessarily \emph{true}), to talk of truth. I should point out that I am nowhere saying that the various virtues exhibited by string theory warrant belief in its absolute truth. What I am suggesting is that they warrant an increase in the credibility of the theory. They make it perfectly \emph{rational} to pursue string theory, and yes, perhaps fund string theory more than its competitors, in spite of the lack of direct experimental support. I know that many string theorists do not think of their theory as `definitely true' but simply as the best available approach. In this article I have attempted to show that this is a perfectly reasonable position to adopt.

I should point out that Cartwright and Frigg are writing from a Lakatosian perspective, according to which research programmes that are able to make novel predictions are considered progressive and those that don't are considered degenerative.  On this account it is not enough to fit a body of evidence, however varied and variegated that body might be. But this tags as degenerative virtually all quantum gravity research, including those programmes that have had `success' in mathematics and other areas, such as computing.  Hence, if the Lakatosian approach has this implication, then I would suggest that the approach itself is at fault: it is too restrictive.

\section{Conclusion}

String theory has not yet been able to make contact with experiments that would give us strong reasons to accept it as the `sure winner' in the race to construct a theory of quantum gravity. However, though experiment can often function as a decisive arbiter in situations where there are several competing theories, there are many more theoretical virtues that play a role in our evaluation of theories. Taking these extra-experimental factors into account, string theory is very virtuous indeed. Not only is it able to unify a whole swathe of old data, and offer the prospect of a consistent theory of quantum gravity (in itself, no mean feat!), it is arguably the most mathematically fertile theory of the past century or so. Though a novel (genuine) physical prediction would perhaps be `worth more' than the combined network of confirming evidence of string theory, until this comes about (within string theory or a competitor) string theory stands a little ahead of the competition. I would go further and say that no direct experiment is likely to ever come about (other than ones that could be explained by multiple approaches), so we can assume that non-experimental factors will have to be relied upon more strongly in our assessments of future research in fundamental physics.

\end{document}